\documentclass[preprint,showpacs,preprintnumbers,amsmath,amssymb, superscriptaddress,longbibliography,nofootinbib]{revtex4-1}

\usepackage{textcomp}
\usepackage{makeidx}
\usepackage{amsmath}
\usepackage{subfigure}
\usepackage{amssymb}
\usepackage{hyperref}
\usepackage{graphicx}
\usepackage{epstopdf}
\usepackage{color}
\usepackage{soul}

\begin{document}

\title{Regular black holes and its thermodynamics in Lovelock gravity.}

\author{Rodrigo Aros}
\email{raros@unab.cl}
\affiliation{Departamento de Ciencias Fisicas, Universidad Andres Bello, Av. Republica 252, Santiago,Chile}

\author{Milko Estrada}
\email{milko.estrada@gmail.com}
\affiliation{Department of Physics, Universidad de Antofagasta, 1240000 Antofagasta, Chile}

\date{\today}

\begin{abstract}
In this work two new families of non-singular or regular black hole solutions are displayed. These black holes behave as de Sitter space near its center and have a well defined AdS asymptotic region for negative cosmological constant. These solutions are constructed on a general ground through the introduction of a finite density of mass/energy. This removes the usual singularity of a black hole and also introduces a new internal geometry. The thermodynamic properties of these solutions are discussed as well.
\end{abstract}


\maketitle

\section{Introduction}

One of the most relevant predictions of General Relativity was the existence of black holes and nowadays there is substantial evidence that this is a usual phenomenon in nature. Now, despite the Schwarzschild black hole solution has been known for over a century it was not until the end of 60's that was shown that, under general {\it classical} conditions, that the formation of a black holes is unavoidable provided the energy (density) within a region of the space surpasses a certain limit. Indeed, the matter in that region collapses until an (event) horizon is formed. However, under the same argument Penrose determined that the final stage of that gravitational collapse gives rise to a singularity as predicted by the existence of Schwarzschild and Kerr solutions. 

One unwanted consequence of the presence of singularities is that they break predictability. Fortunately, it is certain that {\it classical general relativity cannot be valid at all scales} \cite{planckstar1} and thus at Planck scales the description of nature must change drastically. It is precisely this new description which is expected to provide a tamed version of the singularities once {\it quantum effects} are considered. This scenario is supported by results in either String Theory or Loop Quantum Gravity. Results in LQG, for instance, determine that before matter can reach the Planck density, quantum (gravity) fluctuations actually generate enough pressure to counterbalance \emph{weight}. For the physics of a black hole, this implies that the gravitational collapse stops before a singularity can be formed. Furthermore, this can be understood as the formation of a dense central core whose density is of the order of magnitude of the Planck density. These objects are called {\it Planck stars} \cite{planckstar}. Once one has embraced the idea that inside a black hole, instead of a singularity, a dense core exists, one has to propose model for it. The first approximation to do this is to treat the problem as a classical gravitational problem with an energy-momentum density which condenses the quantum effects, in particular the existense of a pseduo repulsive force at the origin. In practice Planck stars can be studied as a geometry which \emph{far away} from the core recovers a standard black hole solution, says Schwarzschild for instance, but whose center, although contains a dense core, can still be treated as a manifold. Moreover, the core of the geometry must approach, in a first approximation, a de Sitter space as such the geodesics diverge mimicking the repulsing force mentioned above. This kind of solutions are called non-singular or {\it regular black holes} and in the context of this work can be considered synonyms to Planck star. 

Historically one of the first regular black hole solutions was found by J. Bardeen \cite{bardeen}. This corresponds to the spherically symmetric space described by
\begin{equation}
ds^2 = -f(r) dt^2 + f(r)^{-1} dr^2 + r^2 d\Omega
\end{equation}
where $f(r)=1- 2m(r)/r$ with $m(r)=Mr^3/(r^2+e^2)^{3/2}$. Evidently $e$ is a regulator, but can also be understood, see \cite{bardeen}, as due to an electric charge density whose electrostatic repulsion prevents the singularity to occur. One can check that $f(r)$ has a zero for $r=r_+>0$, showing the existence of a horizon. By the same token, it is direct to check the absence of singularities. Furthermore,
\begin{equation}
\left. f(r) \right|_{r\approx 0} \approx 1-Kr^2 + O(r^3),
\end{equation}
with $K>0$ showing that this space behaves as a de Sitter space near $r=0$. After \cite{bardeen} others regular black holes have been also studied. See for instance  \cite{planckstar2,7,17,Hayward:2005gi,Sakharov:1966aja,Bambi:2013ufa,BardeendeSitter2,BardeendeSitter3,BardeenRotante,RegularBH1,RegularBH2,RegularBH3,RegularBH4,RegularBH5} and for the higher dimensional case see \cite{BardeendeSitter1,BardeenAdS,GaussBonnetRegular}. As mentioned one must expect an energy density concentrated around $r\approx 0$ which rapidly decays as $r$ grows. This simple idea led, for instance, in \cite{planckstar2} to propose an Gaussian energy density profile, {\it i.e.}, $\rho \propto \exp(-r^2)$ and to Dymnikova in \cite{7,17}  to propose that $\rho \propto \exp(-r^3)$.

\subsection*{Lovelock Gravity}

One of the fundamental aspects of GR, which almost singles it out in three and four dimensions, is that its equations of motion are of second order and thus causality is guarantied. In higher dimensions, $d>4$, however having second order equations of motion is a property of much larger families of theories of gravity. Among them Lovelock gravities have a predominant r\^ole \cite{1,Banados:1993ur}. 

The Lagrangian of a Lovelock gravity in $d$ dimensions is the addition, with arbitrary coefficients, of the lower dimensional topological densities \cite{1,Banados:1993ur} \footnote{In fact, for $d=3,4$ GR is only member of Lovelock gravities}, \emph{i.e.},
\begin{equation}\label{LovelockLagrangian}
L \sqrt{-g} = \sqrt{-g} \sum_{p=1}^N \alpha_p L_p,
\end{equation}
where $N=\frac{d}{2}-1$ for even $d$ or $N=\frac{d-1}{2}$  for odd $d$ and
\begin{equation}\label{Ln}
L_p = \frac{1}{2^p} (d-2p)!\delta^{\mu_1 \ldots \mu_{2p}}_{\nu_1 \ldots \nu_{2p}} R^{\nu_{1} \nu_{2}}_{\hspace{2ex}\mu_{1} \mu_{2}}\ldots R^{\nu_{2p-1} \nu_{2p}}_{\hspace{2ex}\mu_{2p-1} \mu_{2p}},
\end{equation}
where $R^{\alpha \beta}_{\hspace{2ex}\mu\nu}$ is the Riemann tensor and $\delta^{\mu_1\ldots \mu_n}_{\nu_1 \ldots  \nu_n}$ is the generalized $n$-antisymmetric Kronecker delta \cite{Eguchi:1980jx}.  $\{\alpha_p\}$ is a set of arbitrary coupling constants. The normalization in Eq.(\ref{Ln}) is merely a convention. 

The first two terms in this series are $L_1 \propto R $, the Ricci scalar, and $L_2 \propto R^{\alpha\beta}_{\hspace{2ex}\mu\nu}R^{\mu\nu}_{\hspace{2ex}\alpha\beta}-4 R^{\alpha\nu}_{\hspace{2ex}\beta\nu} R^{\beta\mu}_{\hspace{2ex}\alpha\mu} + R^{\alpha\beta}_{\hspace{2ex}\alpha\beta}R^{\mu\nu}_{\hspace{2ex}\mu\nu}$, the Gauss Bonnet density. In addition $L _0 \propto 1$ is introduced to represent a cosmological constant term. In this case the equations of motion are the generalization of the Einstein equations given by
\begin{equation}\label{NewLNGravityEquation}
 \sum_{p} \alpha_{p} \frac{1}{2^p} (d-2p)!\delta^{\mu_1 \ldots \mu_{2p} \mu}_{\nu_1 \ldots \nu_{2p} \nu} R^{\nu_{1} \nu_{2}}_{\hspace{2ex}\mu_{1} \mu_{2}}\ldots R^{\nu_{2p-1} \nu_{2p}}_{\hspace{2ex}\mu_{2p-1} \mu_{2p}} =  G^{\mu}_{ (LL)\hspace{1ex} \nu} = T^{\mu}_{\hspace{1ex} \nu},
\end{equation}
where $G^{\mu \nu}_{(LL)} = \frac{1}{\sqrt{g}}\frac{\delta }{\delta g_{\mu\nu}} (L\sqrt{g})$ and $T^{\mu\nu}$ is the energy momentum tensor of the matter fields. Notice that $\nabla_\mu G^{\mu\nu}_{(LL)} \equiv 0$ is an identity. 

To analyze the potential asymptotic behaviors of the solutions one needs to do a small digression. Let us consider that $\alpha_p=0$ for $p > I$ . It is direct to check that the equations of motion can be rewritten as 
\begin{equation}\label{LovelockLangrangianalternative}
 G^{\mu}_{ (LL)\hspace{1ex} \nu} \propto \delta_{\mu_1 \nu_1 \ldots \mu_{I}\nu_I \nu}^{\alpha_1\beta_1 \ldots \alpha_{I}\beta_I\mu} (R^{\nu_1 \mu_1}_{\hspace{2ex}\alpha_1 \beta_1} + \kappa_1 \delta^{\mu_1 \nu_1}_{\hspace{2ex}\alpha_1 \beta_1})\ldots (R^{\nu_I \mu_I}_{\hspace{2ex}\alpha_I \beta_I} + \kappa_I \delta^{\mu_I \nu_I}_{\hspace{2ex}\alpha_I \alpha_I}). 
\end{equation}
This seems to indicate, as expected, that any Lovelock gravity should have for ground states constant curvature manifolds. To analyze those backgrounds one can introduce the ansatz $ R^{\nu_1 \mu_1}_{\hspace{2ex}\alpha_1 \beta_1}  = x \delta^{\mu_1 \nu_1}_{\hspace{2ex}\alpha_1 \beta_1}$. This maps Eq.(\ref{LovelockLangrangianalternative}) into $G^{\mu}_{\nu} = P_{l}(x) \delta^{\mu}_{\nu}$ where \begin{equation}\label{IndexialEq}
  P_{l}(x) = \sum_{p=0}^I \alpha_p x^p = (x+\kappa_I)\ldots(x+\kappa_1).
\end{equation}
Now, it is direct to demonstrate that, in general, the $\kappa_i$ can be complex numbers, even though $ \forall \alpha_p \in \mathbb{R}$. This does not only restrict the possible constant curvature solutions, and so the potential ground states of the theory, but also severely constraints the space of solutions with a well defined asymptotic region for arbitrary $\{\alpha_p\}$. Indeed, the only allowed behaviors of a ground sate as those which match a constant curvature and, in turn, are related with the zeros of $P_l$ (See Eq.(\ref{IndexialEq})) in the real numbers \footnote{ Finally, although this is a minor issue, for $d \geq 10$ dimensions the expression in Eq.(\ref{LovelockLangrangianalternative}) cannot be constructed out of the Eq.(\ref{LovelockLagrangian}) due to the mathematical impossibility of finding in general the roots of any polynomial for degree 5 or higher.}. For positive or null $\kappa_p$, which correspond to locally AdS or flat, is possible to define an asymptotic region. That region also corresponds to the allowed asymptotic behavior of the solutions of that branch. The case $\kappa_p$ negative stands appart as in this case the ground state is locally dS and there is no asymptotic region. This can be called a dynamical selection of ground sates and simultaneously of asymptotic behaviors. It is worth mentioning that, as noticed in \cite{JulioOliva}, in certain cases the definition of a {\it ground state} can be extended to non-constant curvature spaces. Those cases will be ignored in this work, however.

There are several known black hole solutions of Lovelock gravity in vacuum ($T^{\mu\nu}=0$). See for instance \cite{Banados:1993ur,Crisostomo:2000bb,14,15,Cai:2006pq,Cai:2009de} and reference therein. However, there are not many known solutions in presence of matter fields, see for instance \cite{3}. This is mostly due to the non-linearities of any theory of gravity, which makes difficult, if not impossible, to solve analytically its equations of motion for an arbitrary matter field configuration. Indeed, only highly symmetric configuration can be studied analytically. For the case of our interest, classically regular black holes have been studied in \cite{BardeenAdS,GaussBonnetRegular} within Einstein Gauss Bonnet theories.

In this work two new families of {\it regular black holes} will be displayed.  These solutions share to belong to families of solutions which have a single locally AdS ground state. Moreover, these families have a single well defined asymptotically locally AdS region which approaches the ground state. These two new families correspond to the generalization of the {\it Pure Lovelock} solutions \cite{14,Cai:2006pq,3} and those discussed in \cite{Crisostomo:2000bb} which have a $n$-fold degenerated ground state.

During the next sections, first the general conditions to be satisfied by the mass density will be discussed. Next, it will be obtained the two families of solutions and analyzed their behavior. Finally their thermodynamics will be displayed. 

\section{A well posed mass definition}\label{AWellPosedMassDensity}

In order to be able to solve analytically a black {\it star}, one can impose a highly symmetric geometry. With this in mind one can consider, as a first step, to neglect the presence of angular momenta. In principle, one could consider non spherical transverse sections as long as they are compact constant curvature $(d-2)-$(-sub-)spaces, but for now this only would complicate the analysis. As it is well known in Schwarzschild coordinates a static spherical symmetric geometry can be described by
\begin{equation}\label{22}
ds^2 =-f(r) dt^2+ \frac{dr^2}{f(r)} + r^2 d \Omega^2_{D-2}.
\end{equation}
It is worth to recall that the existence of event horizons is merely determined, due to the geometry, by the zeros of $f(r)$. On the other hand, the energy momentum tensor of a fluid living in this geometry, given the symmetries of the space, must have the form
\begin{equation}\label{TFluid}
    T^{\alpha}_{\hspace{1ex}\beta} =  \textrm{diag}(-\rho, p_r, p_\theta, p_\theta, ...).
\end{equation}
Moreover, it must be satisfied that $\rho=-p_r$ as the lapse function is unitary. Finally, due to $\nabla_{\mu} T^{\mu\nu} =0$, 
\begin{equation}
p_\theta = \frac{r}{d-2} \frac{d}{dr}p_r + p_r. \label{18}
\end{equation}
In general this fluid is usually called an {\bf anisotropic fluid}.

Now we can proceed to analyze the general behavior of the mass density $\rho$. In the next section will be shown that it is convenient, not only to simplify the notation, to define the effective mass function 
\begin{equation}\label{EffectiveMass}
m(r) \propto  -\int_0^r   T^{0}_{\hspace{1ex} 0} r^{d-2} dr =  \int_0^r   \rho(r) r^{d-2} dr.
\end{equation}
In order to have a well posed physical situation it must be satisfied the following conditions; 
\begin{enumerate}
  \item $\rho$ must be a positive due the weak energy condition and a continuous differentiable function to avoid singularities. This implies that $m(r)$ is a positive monotonically increasing function ($m(r)>0$ $\forall r$ and $m(r_1) > m(r_2)$ if $r_1>r_2$) which vanishes at $r=0$.
  \item $\rho$ must have a finite single maximum at $r=0$, the core, ( $\rho(0) > \rho(r)$ $\forall r >0$) and to rapidly decrease away from the core. This yields the condition 
\begin{equation}
 \left. m(r) \right|_{r\approx 0}  \approx K r^{d-1},
\end{equation}
with $K>0$ proportional to $\rho(0)$. The finiteness and snootness of $\rho(0)$ forbid the presence of a curvature singularity at $r = 0$ \cite{planckstar2}. However, it must be noted that this is not enough to ensure a dS behavior near the center of the geometry, and thus additional conditions will be imposed in the next sections. 

\item For the space to have a well defined asymptotic region, such as those to be studied, and to describe a physical object, $\rho(r)$ must be such that $m(r)$ be bounded for $0 < r <\infty$ , \emph{i.e.}, with a well defined limit for $r \rightarrow \infty$. Therefore,
\begin{equation}\label{LimitofM}
\lim_{r \rightarrow \infty} m(r) = {M},
\end{equation}
for ${M}$ some constant. Later it will be shown that ${M}$ is proportional to the total mass of the geometry. This implies that :

\begin{equation}\label{MassDensityFlatering}
\lim_{r \rightarrow \infty} \frac{d}{dr} m(r) = 0.
\end{equation}

\item As mentioned above the idea of a regular black hole is to mimic the exterior of black hole. For this to happen the density $\rho$ must be such that there is a radious $r=r_*$ where is satisfied $m(r_*) \approx M$ and $\frac{d}{dr} m(r_*) \approx 0$.  In general, one can also expected that for large masses that $\ell_P \ll r_{*} \ll r_{+}$ be satisfied. This condition, however, is not satisfied for masses within the range of Plank scales but still the thernodynamics can be studied  \cite{7,17}.  
\end{enumerate}

\section{First family of solutions: Regular Black Holes in Pure Lovelock theory.}
\label{primerafamilia}

As a first step it will be considered a gravitational theory in $d$ dimensions whose Lagrangian is a single term in Eq.(\ref{LovelockLagrangian}) plus a cosmological constant, {\it i.e.},  $L = \alpha_n L_n + \alpha_0 L_0$ where \begin{equation}\label{LastLovelock}
L_n= \delta^{\mu_1 \ldots \mu_{2n}}_{\nu_1 \ldots \nu_{2n}} R^{\nu_{1} \nu_{2}}_{\hspace{2ex}\mu_{1} \mu_{2}}\ldots R^{\nu_{2n-1} \nu_{2n}}_{\hspace{2ex}\mu_{2n-1} \mu_{2n}},
\end{equation}
with $2n < d$. From now on $\alpha_0$, which can be understood as the cosmological constant, will be normalized such that
\begin{equation}
\alpha_0 = \pm \frac{(d-1)(d-2n)}{d} \frac{\alpha_n}{l^{2n}} = - 2\Lambda,
\end{equation}
with $l^2>0$.

The symmetries of the ansatz considered (see Eq.(\ref{22})) leave just one (relevant) equation of motion to be solved. This is given by,
\begin{widetext}
\begin{equation}
\rho(r)r^{d-2} = \alpha_n (d-2n)(d-1)!\left(\pm (d-1)\frac{r^{d-2}}{l^{2n}} + \frac{d}{dr} ( r^{d-2n-1}\left(1-f(r)\right)^n )\right)\label{26}.
\end{equation}
\end{widetext}
The direct integration of Eq.(\ref{26}) defines 
\begin{equation}\label{34}
G_n m(r)   =  r^{d-2n-1}(1-f(r))^n \pm \frac{r^{d-1}}{l^{2n}},
\end{equation}
where
\begin{equation}
   \alpha_n= \frac{1}{\Omega_{d-2} (d-2n)(d-1)! G_n},
\end{equation}
with $G_n$  a constant of units $L^{d-2n}$, where $L$ represents a unit of length. Following the defintion above, 
\begin{equation}\label{mass}
m(r) =  \Omega_{d-2} \int_0^r  \rho(R) R^{d-2} dR.
\end{equation}
is the mass function defined above. For simplicity, in this work, we consider arbitrarily that the constant $G_n$ has a magnitude equal to $1$ (see appendix \ref{Units}) . Finally, by manipulating Eq.(\ref{34}),
\begin{equation} \label{f}
f(r) = 1- \left(\frac{m(r)}{r^{d-2n-1}} \mp  \left(\frac{r^2}{l^2}\right)^n \right) ^{\frac{1}{n}}.
\end{equation}

\subsection{Global analysis} \label{AnalisisGlobalPL}

As mentioned above, any zero of $f(r)$ defines an event horizon in the geometry. This fact significantly simplifies the analysis. Now, to proceed, the cases $\Lambda>0$ and $\Lambda<0$ will be discussed separately.

\subsubsection{\textbf{$\Lambda < 0$ or Negative cosmological constant}} 

The first to notice is that for even $n=2k$, with $k \in \mathbb{N}$, and $\Lambda <0$ ($-$ sign in Eq.(\ref{f}) ) $f(r)$ can take imaginary values. To avoid that it is necessary that
\begin{equation}\label{ToBePositive}
\left(\frac{m(r)}{r^{d-4k-1}} - \left(\frac{r^2}{l^2}\right)^{2k} \right) > 0,
\end{equation}
which occurs only for a certain ranges of $r$.  This rules out the existence of an asymptotic region, defined by $r\rightarrow \infty$, and even an interior region. Notice that there is not an (locally) AdS solution in this case and thus there is no ground state in the spectrum of the solutions. By observing Eq.(\ref{IndexialEq}), this can traced back to the fact the equation $x^{2k} = -1$ has no roots in $\mathbb{R}$.

On the other hand, for odd $n=2k+1$ there are no constraints to be satisfied, and since $m(r)>0$, then there could be a well defined asymptotically AdS region in this case. $f(r)$ can be rewritten as
\begin{equation}\label{f1}
f(r) = 1+ \frac{r^2}{l^2}\left(1-\frac{l^{4k+2}}{\alpha_n(d-2n)(d-1)!}\frac{m(r) }{r^{d-1}}\right)^{\frac{1}{2k+1}}.
\end{equation}

Before to proceed a comment is to be made about the space of solutions. In principle one could have expected to have more than a potential asymptotic behavior in the spectrum of the solutions. However, from equation (\ref{IndexialEq}), one can notice that the allowed asymptotic behaviors are determined by $x^{2k+1} = -1$, whose only real root is $x=-1$. This implies that an asymptotically (locally) AdS behavior of radius $l$ is the only allowed asymptotic behavior in this case. As a consequence, in this work, it will be considered the case where $n$ is odd when $\Lambda<0$ in the Pure Lovelock solution as a representative of the general features.  

Now, assuming $n=2k+1$ one can analyze the structure of the horizons.  The zeros of $f(r)$ can be studied qualitatively from the equation
\begin{equation}\label{ZeroOfFirstSolution}
  \frac{m(r)}{r^{d-2n-1}} = 1 + \frac{r^{2n}}{l^{2n}}.
\end{equation}
By recalling that $m(r)>0$, and in general, one can demonstrate the existence of up to two zeros of $f(r)$ and therefore the presence of up to two horizons. Those radii will be called $r_{-} < r_{+}$. Later it will be identified $r_+$ with the external radius of the black hole horizon. As usual there is an extreme case when both horizons merge ($r_+ \rightarrow r_-$) and thus the space has zero temperature. To address analysis of the different cases requires of numerical analyis which be will carried out later in this work. For now, it is noteworthy that the absence of a horizon, unlike for the usual black hole solution, does not rule out the solution, as no singularities are presented. It only rules out a direct thermodynamic interpretation. 

\subsubsection{\textbf{$\Lambda >0$ or positive cosmological constant}}

For a positive cosmological constant ($\Lambda > 0$) the analogous of Eq.(\ref{ToBePositive}) in Eq.(\ref{f}) is always positive and thus the family of solution contains a well defined ground state. Once again, this could have been foreseen from Eq.(\ref{IndexialEq}) which in this case becomes $x^n =1$ and has always zeros in $\mathbb{R}$. Indeed, for $n=2k+1$, $x=1$ is the only solution in $\mathbb{R}$ and determines that dS spaces is a solution and a ground state in this case. Remarkably for $n=2k$, $x=\pm 1$ are both solutions, and therefore, for $\Lambda>0$ it is also possible to construct solutions with an AdS asymptotia. This produces a solution whose behavior mimics the previous case with $\Lambda <0$ and $n=2k+1$ mentioned above, thus no further attention is necessary in this case. 
Before to proceed, let us have a digression about the positive root which is connected with a dS ground state. The horizons can be analyzed by studying graphically the relation
\begin{equation}
  \frac{m(r)}{r^{d-2n-1}} = 1 - \frac{r^{2n}}{l^{2n}}.
\end{equation}
Since $m(r)>0$ extrictly, it is direct to check the existence of up to three horizons in this case. These can be called $r_{-}<r_{+}<r_{++}$, respectively. One utmost relevant consideration to be made is that $r_{++}$ defines a cosmological horizon but also the outer spatial boundary of the space, ruling out the existence of an asymptotic region, in the sense of a $r\rightarrow \infty$ region, for the space. This condition separes this case from the asymtotically AdS case fundamentally. Finally, in the in-between region $r_{-} < r < r_{+}$ can be interpreted as Kantowski-Sachs cosmology.

\subsection{Internal Geometry} \label{AsymptoticalConditions}

The previous construction was performed without much concern of the details of $\rho(r)$. One can, in fact, use any of the densities mentioned above \cite{bardeen,planckstar1,7,planckstar2}. Fortunately, the same can done for small $r$. As mentioned above, $m(r)$ must be such that
\begin{equation} \label{SmoothCondition}
    \left. m(r) \right|_{r\approx 0} \approx K \cdot r^{d-1},
\end{equation}
with $K>0$ a constant, and thus equation (\ref{f}) for $r \to 0$ behaves as:
\begin{equation}
\left. f(r) \right|_{r\approx 0}  \approx 1 - \left (K \mp \frac{1}{l^{2n}}\right)^{\frac{1}{n}} r^2 = 1 - \frac{r^2}{l_{eff}^2}. \label{PureLovelockRegularBH}
\end{equation}
Moreover, one can check that the curvature invariants are finite and thus the geometry is smooth everywhere. Now, in order to have a non-collapsing region near $r=0$, {\it i.e.}, to describe a regular black hole \cite{planckstar1}, it is necessary to impose that at  $r \approx 0$, the geometry approximates a locally dS space \cite{BardeenAdS}. This last condition determines that $K > \frac{1}{l^{2n}}$, which in turn corresponds to a constraint on the mass density $\rho(r)|_{r=0}$. It is worth to mention that regular black holes with an effective AdS behavior near their origins have been studied in \cite{BardeenAdS}.

\subsection{The non spherical symmetric solutions}

As mentioned above it is possible to explore solutions with null and negative constant curvature transverse sections, instead of only the spherical symmetry. The geometry to be considered in this case is given by 
\begin{equation}\label{Schw-Gene}
    ds^2 = -f(r) dt^2 + f(r)^{-1} dr^2 + r^2 d\Sigma_{\gamma}
\end{equation}
Here it was replaced the spherical transverse geometry in Eq.(\ref{22}) by a closed ($d-2$-) manifold of constant curvature $\gamma$.  See \cite{Aros:2000ij} for a discussion, In this case the solution is given by
\[
f(r) = \gamma- \left(\frac{m(r)}{r^{d-2n-1}} \mp  \left(\frac{r^2}{l^2}\right)^n \right) ^{\frac{1}{n}}.
\]
It can be noticed that the analysis of the solution is analogous to the spherical symmetry. This is mostly due to the presence of the $n$-root of the same expression, $\left(\frac{m(r)}{r^{d-2n-1}} \mp  \left(\frac{r^2}{l^2}\right)^n \right) ^{\frac{1}{n}}$, whose analysis is independent of the value of $\gamma$. Therefore, the families of solutions with asymotically locally AdS region are defined accoordingly. For instance, the solution exists only for $n=2k+1$ and $\Lambda <0$ or for one of the roots of $\Lambda >0$, but the analyisis of the horizon is the same for both cases. After recalling that $m(r) > 0$ must be strictly satisfied, one can notice that for $\gamma=0$ $f(r)$ has always two roots, $r=0$ and $r_+ \neq 0$. One can also notice that $r=0$ defines an irrelevant horizon for the discussion in this work and thus only $r=r_+$ defines a relevant horizon. For $\gamma=-1$ the situation is slightly different as there is always a single horizon. This also will be $r_+$, as usual.

\section{Second family of solutions: Regular black holes with $n$-fold degenerated ground state} \label{segundafamilia}

As mentioned above in general the Lovelock gravity might have more than a single constant curvature ground state, which makes those ground states unstable under dynamical evolution. One way to avoid this, observe Eq.(\ref{IndexialEq}), is by choosing the ${\alpha_p}$ \footnote{Roughly speaking one can avoid this to occurs if the equation of motions have the form
\begin{equation}\label{BHS}
  \frac{\delta }{\delta g_{\mu\nu}} L\sqrt{g} \sim ((R + l^{-2})^n)^{\mu\nu}.
\end{equation}
This can be obtained provided the parameters $\alpha_p$ of Lovelock Lagrangian equation (\ref{LovelockLagrangian}) are given by \cite{Crisostomo:2000bb}
\begin{equation}
\alpha^n_p = \left\{\begin{array}{cl}
                    \frac{\alpha_n}{d-2p}\binom{n}{p} &\textrm{ for } 0 \leq p \leq n\\
                    0 &\textrm{ for } n < p \leq N
                  \end{array}
\right.,
\end{equation}
where $\alpha_n$ is a global coupling constant.}
such that $P_l(x)$ becomes $P_{l}(x) = (x \pm l^{-2})^n$ creating $n-$fold degenerated ground state of constant curvature $\pm l^{-2}$. From now on only the {\it negative cosmological constant} case will be discussed as this the case where a genuine asymptotic region exists. Still some comment on the {\it positive cosmological constant} case will be done when they are straightforward.  

The static black hole solutions of these theories with {\it negative cosmological constant} have been studied in \cite{Crisostomo:2000bb,Aros:2000ij}. Remarkably, their generalization to the case under discussion is straightforward. The geometry to be consider in this case is given by Eq.(\ref{Schw-Gene})
\begin{equation}\label{Schw-Gene2}
    ds^2 = -f(r) dt^2 + f(r)^{-1} dr^2 + r^2 d\Sigma_{\gamma}
\end{equation}
where the spherical transverse geometry in Eq.(\ref{22}) has been replaced by closed ($d-1$-) manifold of constant curvature $\gamma$, 

As for the previous solution the symmetries are strong enough to just left one equation to be solved. This is given by 
\begin{equation}\label{BHSSolution}
  \alpha_n (d-2n)(d-1)! \frac{d}{dr}\left(r^{d-1}\left[\frac{\gamma - f(r)}{r^2} + \frac{1}{l^2}\right]^{n}\right) = \rho(r) r^{d-2}.
\end{equation}
 After some manipulations, 
\begin{equation}\label{BHSGsolution}
  f(r) = \gamma + \frac{r^2}{l^2} - \left(\frac{m(r)}{r^{d-2n-1}}\right)^{1/n},
\end{equation}
where $m(r)$ can be in principle any function that satisfies the criteria introduced in section \ref{AWellPosedMassDensity} . It can be noticed, from Eq.(\ref{BHSGsolution}), that, besides $d > 2n$, there is no restriction on the power $n$ or the dimension $d$.

\subsection{Horizons}
As previously, the structure of zeros of $f(r)$, and thus of the horizons, can be studied qualitatively from analyzing graphically
\[
\gamma + \frac{r^2}{l^2}  =  \left(\frac{m(r)}{r^{d-2n-1}}\right)^{1/n}.
\]
One first must recall that $m(r)>0$ which is a difference with case discussed in \cite{Aros:2000ij}. It is straightforward that the roots of $f(r)$ depend on $\gamma = 0,\pm 1$. For $\gamma=1$, $f(r)$ has up to two different roots, to be called $r_- < r_+$, which can coalesce in an extreme case, As before, due to the absence of singularities, the solutions with no horizons cannot be reuled out. For $\gamma=0$, $r=0$ is a root and there is always a second root which will be called $r_+$, For $\gamma=-1$ there is always a single root, to be called $r_+$ as well.

\subsection{Limits of this solution.} \label{geometriabhs}

Recalling section \ref{AWellPosedMassDensity} one has the conditions to be satisfied by $m(r)$, which in this case imply 
\begin{equation}
\lim_{r \to \infty} f(r) \sim \gamma + \frac{r^2}{l^2} - \left(\frac{ {M}}{r^{d-2n-1}}\right)^{1/n}  \sim \gamma + \frac{r^2}{l^2}.
\end{equation}
and thus these solutions approach the locally AdS ground state for large $r$, as expected. In the same fashion, 
as mentioned in section \ref{AWellPosedMassDensity}, as $r\rightarrow 0$ is satisfied that $m(r) \approx K r^{d-1}$. This determines
\begin{equation}
\lim_{r \rightarrow 0} f(r) = \gamma - r^2 \left( K^{1/n} - \frac{1}{l^2}  \right). \label{regularbh}
\end{equation}
It can be noticed that this constrains $K > l^{-2n}$ in order to have a nearly (locally) dS geometry at the origin as required to model a regular black hole. .

\section{The thermodynamics before the thermodynamics}

After studying the properties of these solutions in a general framework one can proceed to analyze the thermodynamics of the black holes in the spectrum of solutions. As mentioned previously, the central concern in this work are only spaces with a well defined asymptotically locally AdS region, but still some general comments will be made as well. For instance, although the number of horizons, and their properties, depends on $m(r)$, there is always a  particular horizon which can be unambiguously cast as the horizon of the black hole. The associate zero will be denoted $r_+$. This is even independent of the existence, for $\Lambda > 0$, of an even outer (cosmological) horizon on the geometry. 

It is utmost relevant to stress that the thermodynamics of black holes whose ground states are locally AdS or flat spaces differs of the thermodynamics of those whose graound state is a locally dS space. This is due to the obvious fact, mentioned already, that while for the latter there exists a well defined asymptotic region, for later there is no \textit{asymptotic} region at all. It is well known that temperature and entropy can be defined independently of the existence of asymptotic region. However, the existence of conserved charges cannot, in particular the mass. In this way, a standard first law of thermodynamics, {\it i.e.} $TdS = dE + \ldots$, can only be developed for spaces with a well defined asymptotic region. The lack of an asymptotic region, or the presence of a cosmological horizon, forbids such a kind of relations to exist. Moreover, in \cite{Aros:2008ef} was established that the presence of a black hole in geometry which also has a cosmological horizon can be interpreted as non-equilibrium thermodynamic system. That system,  through the evaporation processes of both horizons, evolves into a de Sitter space. One can conjecture, for the solutions above, that the same can happen. That analysis, however, will be carried out elsewhere. 

In what follows it will be analyzed the thermodynamics for asymptotically locally AdS spaces whose have a single well defined locally AdS ground state. 

\section{Thermodynamics of asymptotically AdS solutions} \label{TermodinamicaAdS}

In this case the thermodynamics can be obtained following standard Wald's prescriptions based on the adiabatic change of the Noether charges in the region $]r_+ , \infty[ $ \cite{Lee:1990nz}. First, the mass of a solution can be determined in terms of the Noether charge associated with a timelike Killing vector in the asymptotic region. In this case above $\partial_{t}$ in Eq.(\ref{22}) is that Killing vector. In principle a second contribution to the Noether charge coming from action principle of the matters fields should be considered. However, and also in general, that second part of Noether charge usually becomes negligible at the asymptotic region. In the case at hand those additional parts of Noether charge will be assumed that do not contribute to the asymptotic value of the Noether charge.

\subsection{Analysis of the First Family of solutions}

\subsubsection{\textbf{Charges}}

In this particular case, after a straightforward computation, the Noether charge is given in terms of $f(r) = 1-g(r)^{\frac{1}{n}}$, with 
\begin{equation}
g(r) = \frac{m(r)}{r^{d-2n-1}} - \frac{r^{2n}}{l^{2n}},
\end{equation}
and $n=2k+1$, is given by
\begin{equation}\label{Mass}
\lim_{r\rightarrow \infty} Q(\partial_t) =  \lim_{r\rightarrow \infty} \left( - \alpha_n (d-2)!\left( \frac{dg(r)}{dr} r^{d-2n}\right) \right) \int d\Omega_{d-2} + \mbox{Reg}.
\end{equation}
Here $\mbox{Reg}$ is a regulator that removes the divergences associated with the AdS volume \footnote{This can be done by any of the known methods in the literature. For a discussion in terms of holographic regularization see \cite{Skenderis:2002wp} and for Kounterterms and its equivalence see \cite{Kofinas:2008ub,Olea:2006vd}}.  It can be noticed that to evaluate the total mass is only necessary to know the asymptotic value of $m(r)$. Indeed, if this value is fixed such that $\displaystyle \lim_{r\rightarrow \infty} m(r) = {M}$, then the final result, after regularization following \cite{Kofinas:2007ns}, is
\begin{equation}
\lim_{r\rightarrow \infty} \left. Q(\partial_t) \right|_{\textrm{Reg}} = \tilde{M} + E_0.
\end{equation}
where
\begin{equation}
    \tilde{M} =\frac{(d-2n-1)}{(d-2n)(d-1)} \cdot {M}. \label{MasaTotal}
\end{equation}
Here $E_0 =0$ for even dimensions and the AdS vacuum energy for odd dimensions \cite{Mora:2004rx}\footnote{The arise of a vacuum energy is a feature of odd dimensional asymptotically locally AdS spaces.}. This confirms that the mass parameter $M$ indeed is proportional to the mass/energy of the solution.

\subsubsection{\textbf{Temperature}}
Given the form of the metric in Schwarzschild coordinates, one can follow the standard approach to obtain the expression of the temperature. This is given by
\begin{widetext}
\begin{eqnarray}
  T &=& \frac{1}{4\pi} \left.\frac{df}{dr}\right|_{r=r_+} \nonumber\\
    &=& \frac{1}{4\pi n}\left( \frac{d-2n-1}{r_+} + (d-1)\frac{r_+^{2n-1}}{l^{2n}}- G_n \frac{dm(r)/dr|_{r=r_+}}{r_+^{d-2n-1}} \right)\label{GenericTemperature}\\
    &=& \frac{1}{4\pi n}\left( \frac{d-2n-1}{r_+} + (d-1)\frac{r_+^{2n-1}}{l^{2n}}- G_n \rho(r_{+}) r_+^{2n-1} \right) \nonumber
\end{eqnarray}
\end{widetext}
One can notice that the last term, which is always finite, is responsible for the vanishing of the temperature for certain $r_+$. However, until a particular $\rho(r)$ is given there is no form to determine analytically any further feature of $T(r_+)$. In general, one can predict the existence of two inflection point, as $T(r_+)$ can have a local maximum and minimum for certain values of $r_+$. If so, then there is a range in $[r_+]$ where three different $r_+$ yield the same temperature. As these two inflection points merge the temperature becomes a strictly increasing function. See below for a particular model. 

\subsubsection{\textbf{Entropy}}

The form of the space together with the action principle for gravity allows to define a canonical ensemble for the thermodynamics of these solutions.  By following the standard procedure defined by Wald, the first law of thermodynamics will emerged in its usual form
\begin{equation}
   T dS + \sum_i \Phi_i dq^i = dE,
\end{equation}
with $E=M$ the mass of the black hole and $q^i$ are the additional charges due to the presence of the matter fields and $\Phi_i$ their corresponding potentials. Therefore, to address beyond the mass and the temperature is necessary to fix additional thermodynamic variables, in this case at the horizon, in order to obtain the first law of thermodynamics. Unfortunately to do that analysis one needs to have an action principle for the matter fields. Nonetheless, due its pure geometrical original, the entropy can be computed following the approach proposed by Wald \cite{Wald:1993nt}. The final result in $d$ dimensions is given by
\begin{equation}\label{EntropyFE}
  S = \alpha_n n (d-2)! r_{+}^{d-2n} \Omega_{d-2} + S_0,
\end{equation}
where $S_0$ is a finite term independent of the value of $r_+$ and due to the topological pedigree of the regulator mentioned above \cite{Kofinas:2008ub}. It must be stressed that this expression is totally generic and functional independent of $m(r)$. The value of $r_+$ is not independent of the function $m(r)$, though.  It can be noticed as well that this entropy does not follow an area law by a power $r^{2n-2}_+$ but still defines an increasing function of $r_+$. It is worth to stress that for the particular case $n=1$ the area law is well defined.

To pursue any further into an analytic analysis of the thermodynamics properties, for instance the evaporation of these black holes, would require to define the temperature as an function of the mass and additional chargess of the solution, which in turn needs at least to consider a $\rho(r)$ in particular. Unfortunately, this is not enough in general to obtain close expressions. Because of that in the next sections will be considered a particular case which will be analyzed through numerical methods. 

\subsection{Analysis of the Second Family of Solutions}

\subsubsection{\textbf{Charges}}

As before, mass can be computed by computing the Noether charge associated with $\partial_t$. After regularization (one can follow \cite{Kofinas:2008ub}), is given
\begin{equation}\label{MassBHS}
  Q(\partial_t)^{\infty} = \tilde{M} + E_0
\end{equation}
where  $E_0 =0$ for even dimensions and corresponds to the vacuum energy for odd dimensions. This is in complete analogy with \cite{Crisostomo:2000bb,Aros:2000ij} but for the Chern Simons case. The difference with results in \cite{Crisostomo:2000bb,Aros:2000ij} arise due to the location of the horizons.

\subsubsection{\textbf{Temperature}}
In complete analogy with the previous case the temperature can be computed. For the outer horizon this yields
\begin{widetext}
\begin{eqnarray}
   T &=& \frac{1}{4\pi n}\left( \frac{d-2n-1}{r_+} + (d-1) \frac{r_+}{l^2} - \left( 1 + \frac{r_+^2}{l^2} \right) \left(\frac{1}{m(r)} \frac{dm(r)}{dr}\right)_{r=r_+} \right) \label{BHSGenericTemperature}\\
    &=& \frac{1}{4\pi n}\left( \frac{d-2n-1}{r_+} + (d-1) \frac{r_+}{l^2} - \left( 1 + \frac{r_+^2}{l^2} \right)^{1-n} r_+^{2n-1} G_n \rho(r_+)\right) \nonumber
\end{eqnarray}
\end{widetext}
Notice that the last term is a correction to the expression obtained in \cite{Crisostomo:2000bb} due to the finite density of mass. This correction is responsible for the vanishing of the temperature for a value of $r_+$.

\subsubsection{\textbf{Entropy}}
In general the entropy can be computed as well in this case. This yields
\begin{widetext}
\begin{equation}\label{BHSEntropy}
  S = \alpha_n (d-2)! \left(\frac{r_+}{l}\right)^{d-2}\Sigma_\gamma \left[ \frac{n}{d-2} F\left([1-n,1-d/2],[2-d/2],-\frac{\gamma l^2 + r^2_+}{r^2_+}\right)\right] + S_0,
\end{equation}
\end{widetext}
where $\Sigma_\gamma$ is the area of the transverse section. It is matter of fact that equation (\ref{BHSEntropy}) is same expression obtained for the black hole solutions in \cite{Aros:2000ij} and the differences are due to the value of $r_+$ for a given value of $\tilde{M}$. $S_0$ is due to topological term \cite{Kofinas:2008ub} added to regularize the action principle \cite{Aros:2005by}.

For large $r_+$ this entropy always approach an area law. It can be noticed as well that for $n=1$ the usual area law is recovered. This also occurs for $\gamma=0$ for any value of $n$ or $d$. It is direct to check that for $\gamma=1$ and $\gamma=0$ these functions are monotonically increasing functions of $r_+$. Far more interesting is the fact that $S_{\gamma=-1}(r_{+})$ can be negative which imposes servere constrains in the space of parameters. As previously, $r_+$ cannot be determined analytically and so to proceed a numerical approach would be required. This will be also considered for a next work.

\section{Planck energy density}\label{planckstar}

So far it has been shown that an anisotropic fluids is a suitable model to describe a non-singular black hole. Still, we have not proposed any form for $\rho(r)$ or $m(r)$. Now we will propose a $d$ dimensional generalization of Hayward density \cite{Hayward:2005gi}. For the $4D$ case and Einstein Hilbert theory this model of density was used in references \cite{planckstar1, planckstar} for describe Planck Stars. Our generalization is given by: 
\begin{equation}\label{FiniteMass}
\rho(r) = \frac{d-1}{\Omega_{d-2}} \frac{Q^{d-2}M^2}{(Q^{d-2}M+r^{d-1})^2}
 \textrm{  which yields  } m(r)=\frac{M r^{d-1}}{Q^{d-2}M+r^{d-1}}.
\end{equation}
Here $Q$ is introduced as regulator to avoid the presence of a singularity. It must be emphasized that $Q$ has units of $\ell_p^{d/(d-2)}$, such that the energy density is of the order of a Planck units near from origin and such that $\rho(r)$ be able to satisfy the conditions in section \ref{AWellPosedMassDensity} and thus to describe a regular black hole density \cite{planckstar1}. For instance, one can notice that
\begin{equation}\label{masaydensidadcercaorigenplanck}
\rho_{\textrm{max}} = \rho(0) = \frac{(d-1)}{\Omega_{d-2}Q^{d-2}} \textrm{  and  }  \left. m(r) \right|_{r\approx 0} \approx \frac{r^{d-1}}{Q^{d-2}},
\end{equation}
confirming that the energy density near of origin is of the order of Planck density.

\subsection{Structure of Solutions}
With the mass density in Eq.(\ref{FiniteMass}) is direct, by simple substitution, to obtain the form of the solutions. Moreover, for $r\approx 0$, both solutions behaves as de Sitter spaces with the effective cosmological constant is given by $\Lambda = (d-1)(d-2)l_{eff}^{-2}/2 > 0$. $l_{eff}^2$ differs for both cases, however. 

\subsubsection{{\bf First family of solutions:}}

In the case $f(r)$ is given by
\begin{equation} \label{familia1deSitter}
    f(r)=1 - \left( \left(\frac{M }{Q^{d-2}M+r^{d-1}}\right) - \frac{1}{l^{2n}} \right)^{\frac{1}{n}} r^2,
\end{equation}
which near $r\approx 0$ behaves as 
\begin{equation}
\left. f(r) \right|_{r \approx 0} \approx 1 - \frac{r^2}{l^2_{eff}},
\end{equation}
a de Sitter space with  
\begin{equation*}
    \frac{1}{l_{eff}^2} = \Big ( \frac{1}{Q^{d-2}}-\frac{1}{l^{2n}} \Big )^{\frac{1}{n}},
\end{equation*}
where it is required that $l^{2n}>Q^{d-2}$.

\subsubsection{{\bf Second family of solutions:}}

In this case,
\begin{equation}
f(r)=1- \left(-\frac{1}{l^2} + \left(\frac{M}{Q^{d-2}M+r^{d-1}}\right)^{\frac{1}{n}}\right)r^2, \label{f2}
\end{equation}
which also satisfies $\left. f(r) \right|_{r\approx 0}  \approx 1- (r/l_{eff})^2 $, in the case with
\begin{equation}
\frac{1}{l_{eff}^2} = - \frac{1}{l^2} + \frac{1}{Q^{\frac{d-2}{n}}},
\end{equation} 
where again must be fulfilled that $l^{2n}>Q^{d-2}$.

\subsection{Horizons}
\begin{figure}
\centering
\subfigure[$f(r)$ for $n=1,d=5$.]{\includegraphics[width=80mm]{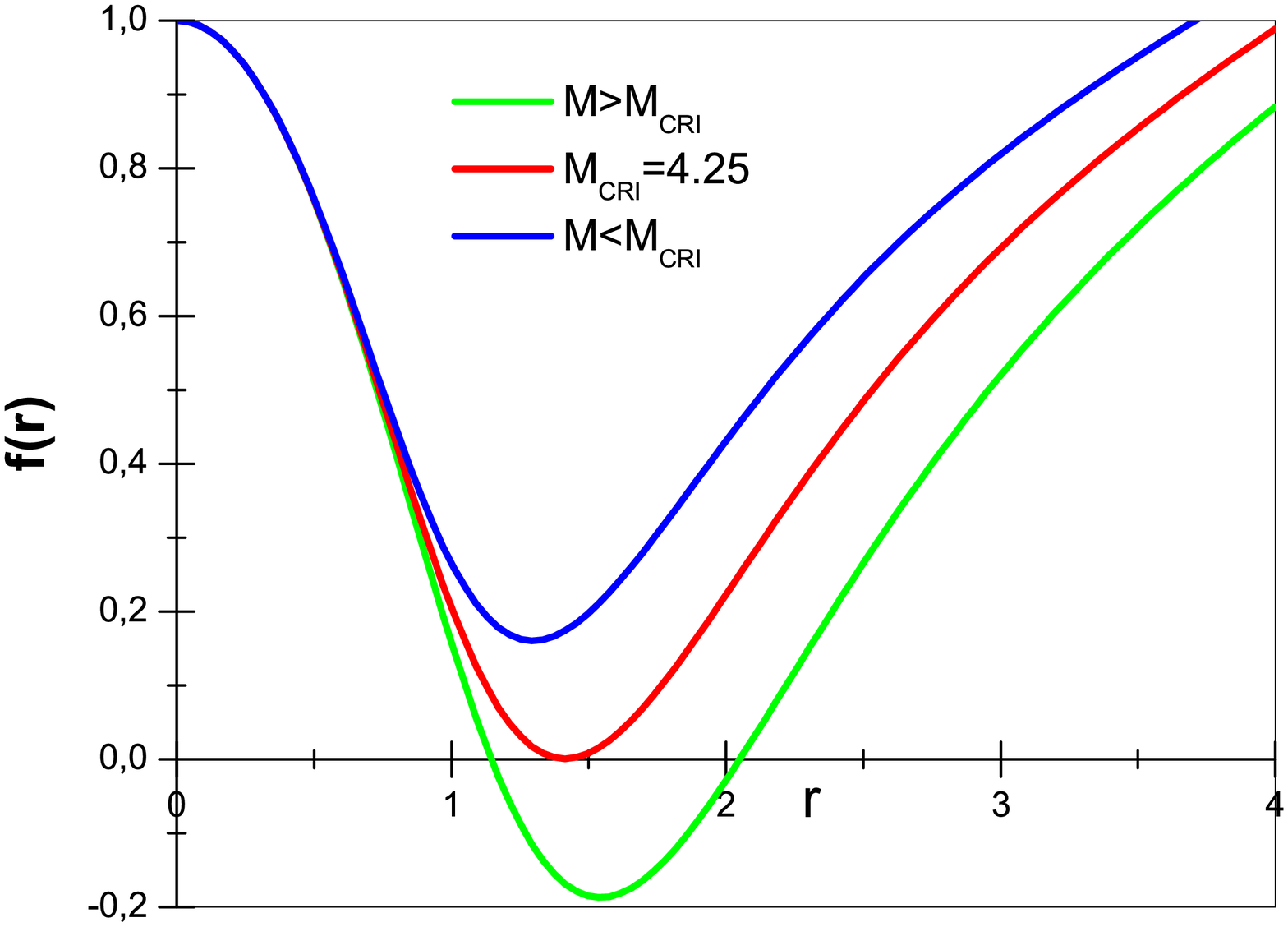}} 
\subfigure[$f(r)$ for $n=3,d=8$.]{\includegraphics[width=80mm]{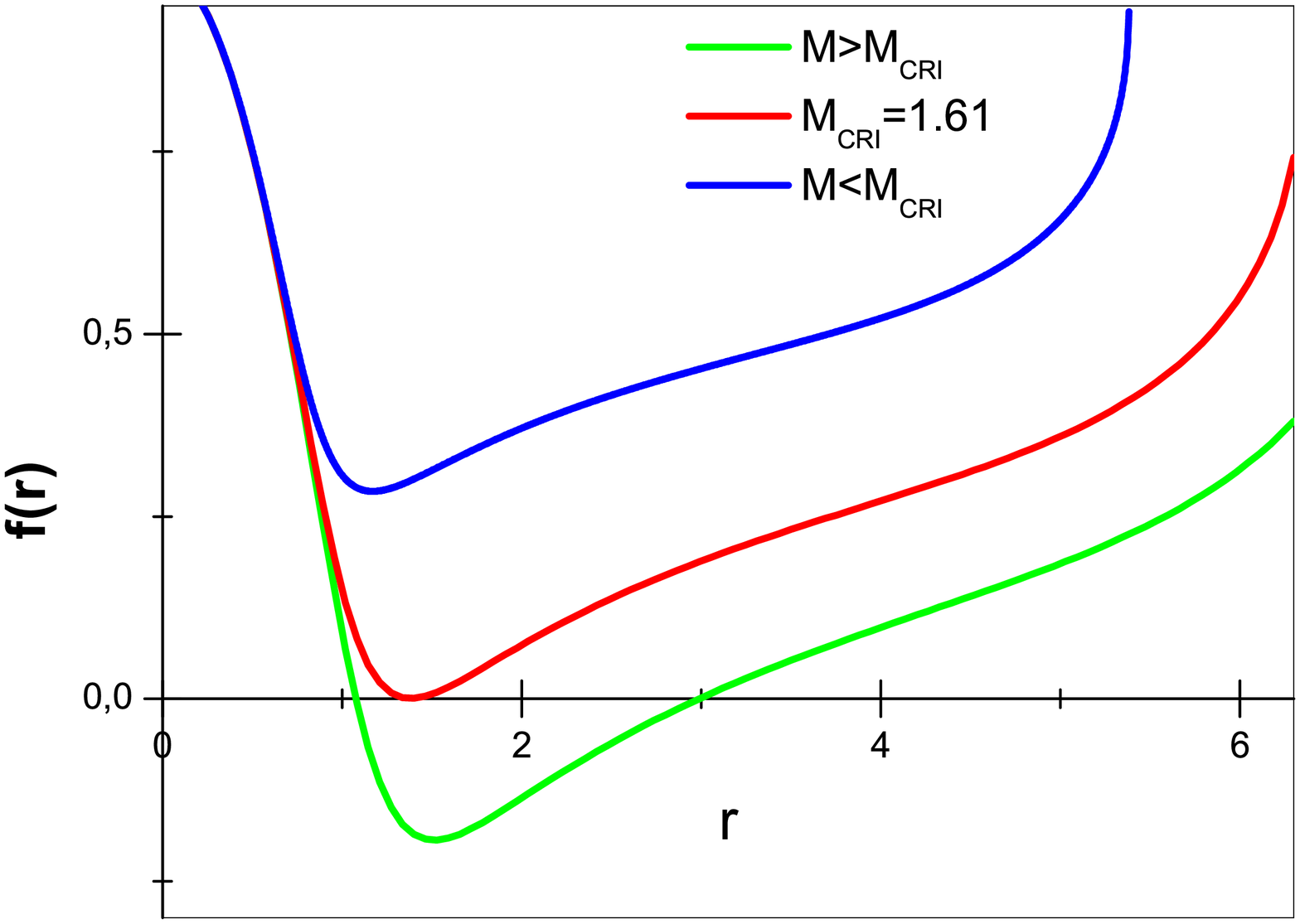}} 
\caption{$f(r)$ for first family of solutions with toy values of $Q=1$ and $l=8$ .}
\label{horizontesPL}
\end{figure}

\begin{figure}
\centering
\subfigure[$f(r)$ for $n=2,d=7$.]{\includegraphics[width=80mm]{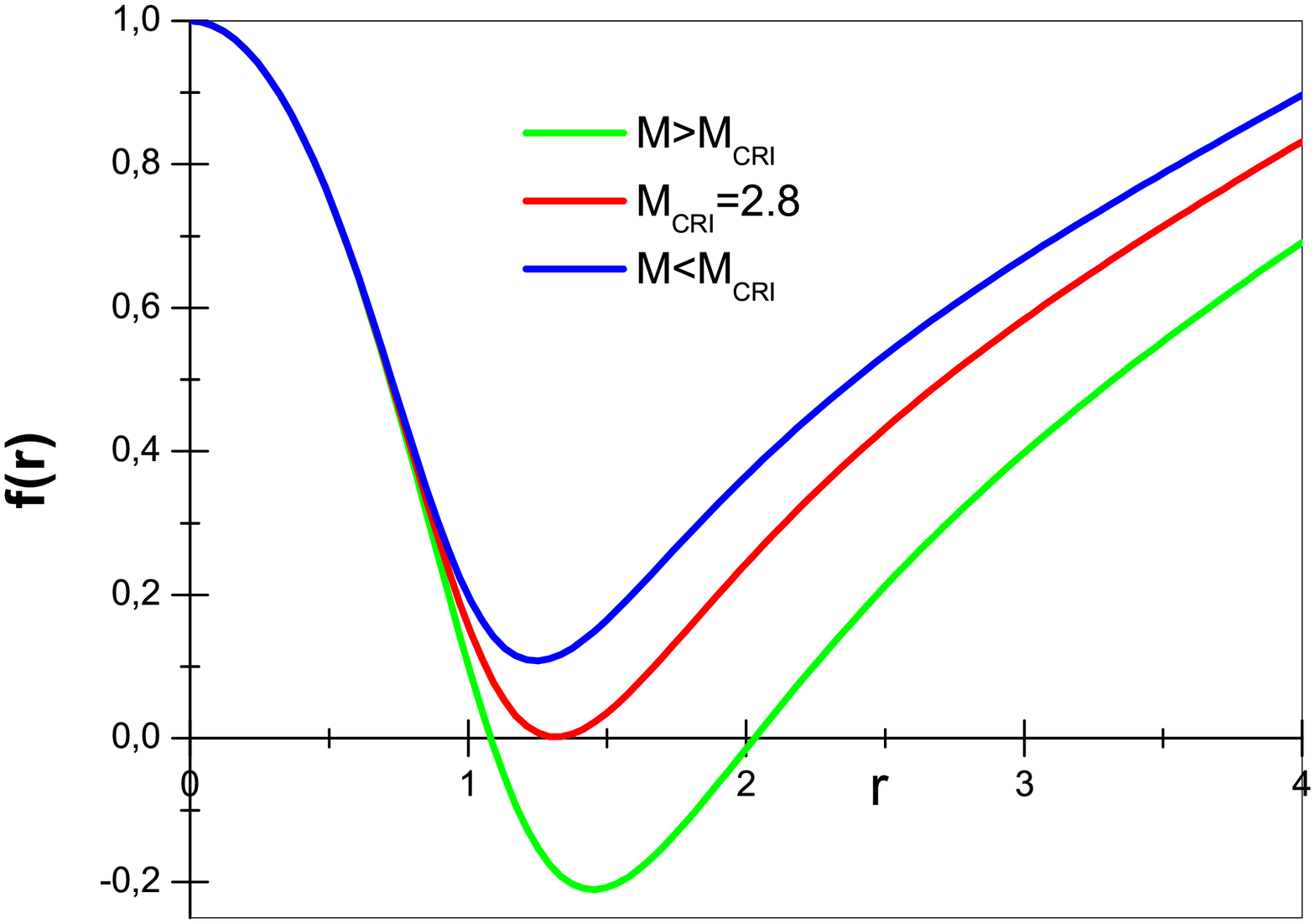}} 
\subfigure[$f(r)$ for $n=3,d=8$.]{\includegraphics[width=80mm]{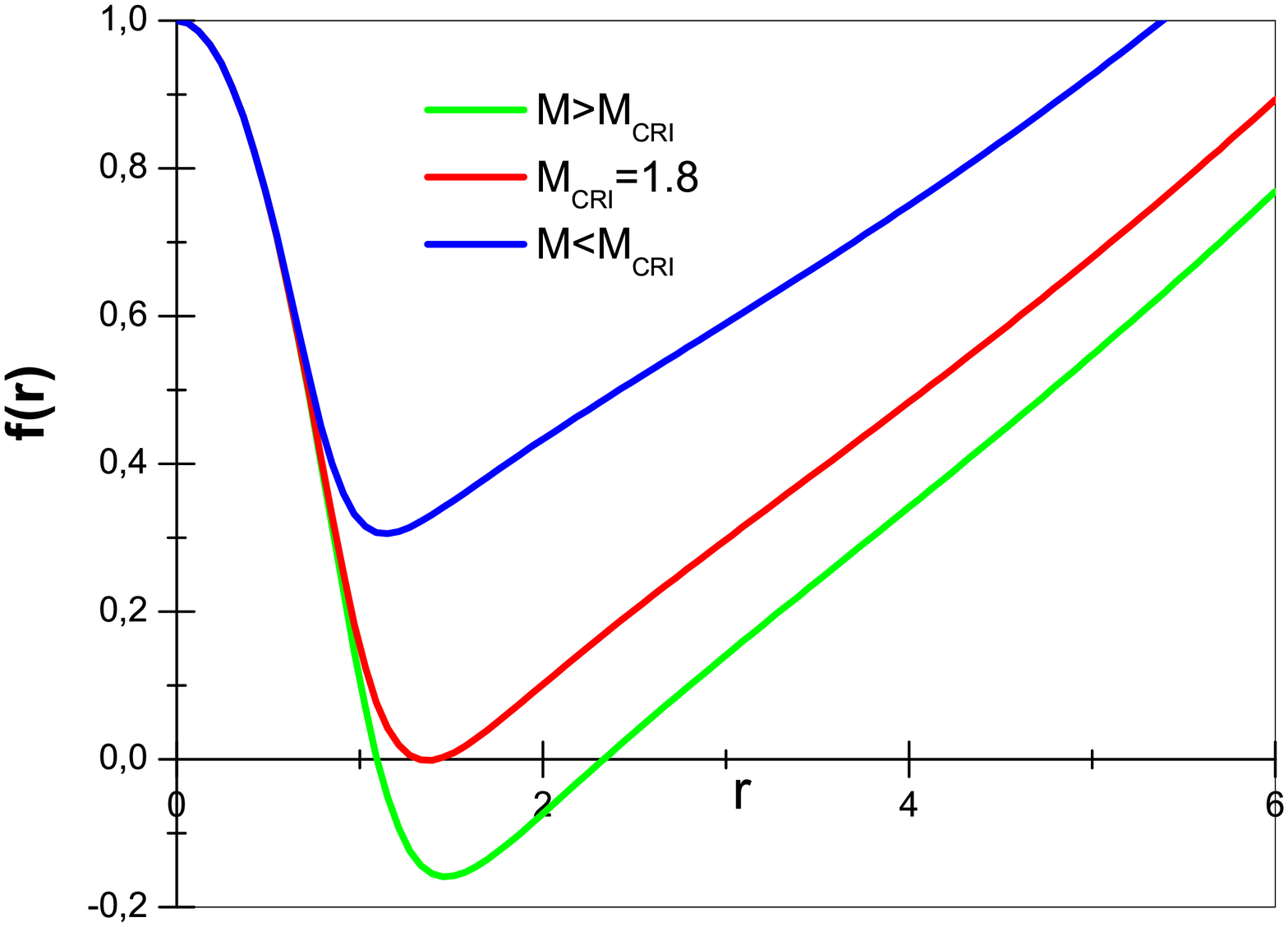}} 
\caption{$f(r)$ for second family of solutions with toy values of $Q=1$ and $l=8$ .}
\label{horizontesBHS}
\end{figure}

We see the behavior of the function $f(r)$ for the first family of solutions in Figure \ref{horizontesPL} and for the second family of solutions in figure \ref{horizontesBHS}. The most relevant observation corresponds to the appearance of mass value where the solution evolves from a black hole into a {\it regular}  star in the sense of a very dense object without a horizon. In this transition the temperature vanishes as mentioned above and can be seen in Figs. \ref{horizontesPL} and \ref{horizontesBHS}. As mentioned previously, that transition for non-regular black hole is meaningless, or nonphysical, as would produce a naked singularity. Therefore, for both solutions it can be distinguished three cases: 
\begin{itemize}
\item for $M<M_{cri}$ $f(r)$ has non zeros in the real numbers and thus the space is not a black hole. Moreover, due to the absence of singularities the space has a causal structure similar to a Minkowski space. 
\item for $M = M_{cri}$ $f(r)$ has a double zero. This represents a regular extreme black hole with degenerate Killing horizon.
\item for $M>M_{cri}$ $f(r)$ has two zeros. This represents a regular black hole with both outer and inner horizons.
\end{itemize}

From now it will explored only the region $M>M_{cri}$. 

\subsection{Temperature}
\begin{figure}
\centering
\subfigure[$T^o$ for $n=1,d=4,5,6$.]{\includegraphics[width=80mm]{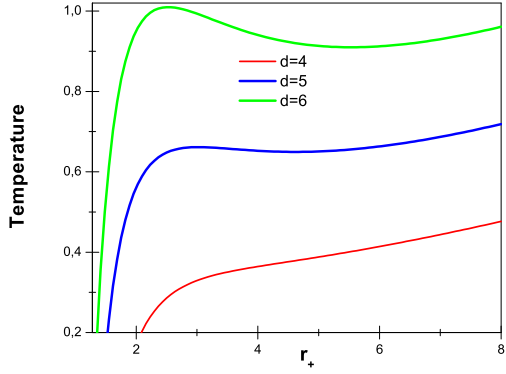}} 
\subfigure[$T^o$ for $n=3,d=8,9,10$.]{\includegraphics[width=80mm]{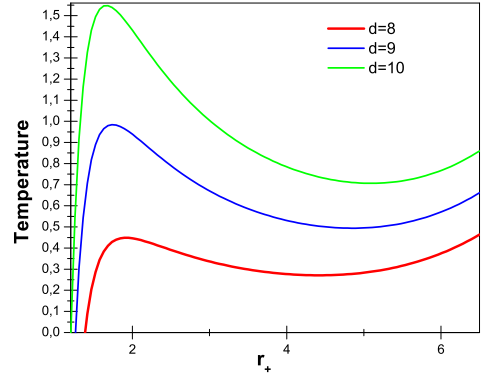}} 
\caption{$T^o$. Figure a) represents both families of solutions, and figure b) represents to the first family of solutions, both with toy values of $Q=1$ and $l=8$.}
\label{TemperaturaPL}
\end{figure}

\begin{figure}
\centering
\subfigure[$T^o$ for $n=2,d=6,7,8$.]{\includegraphics[width=80mm]{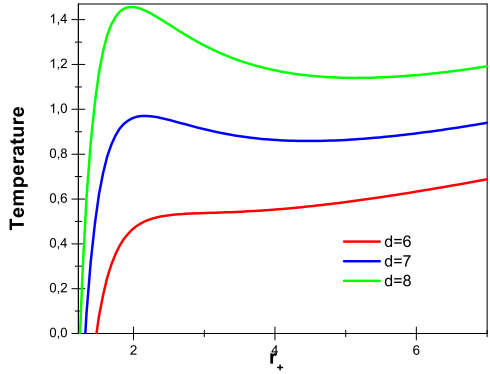}}
\subfigure[$T^o$ for $n=3,d=8,9,10$.]{\includegraphics[width=80mm]{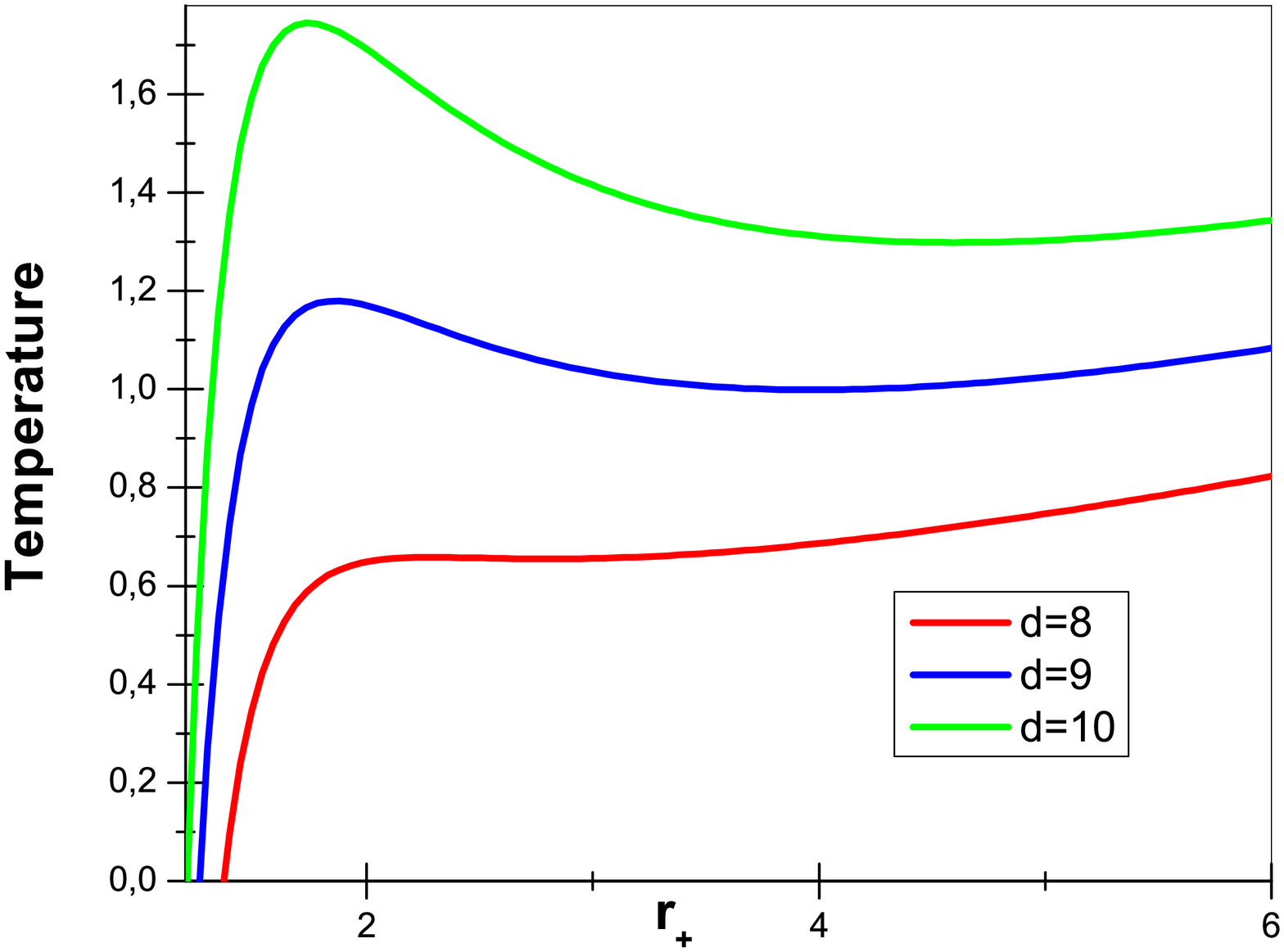}}
\caption{$T^o$ for second family of solutions with toy values of $Q=1$ and $l=8$ .}
\label{TemperaturaBHS}
\end{figure}

Figures \ref{TemperaturaPL} and \ref{TemperaturaBHS} display examples of the behavior of the temperatures, respectively. Figure \ref{TemperaturaPL}(a) actually corresponds to both families of solutions since $n=1$.

In general, one can notice that the temperature function has both a local maximum and a local minimum, says at $r_+=r_*$ and $r_+=r_{**}$ respectively. As $\frac{dT}{dr_+}$ vanishes in those points, the heat capacity $C_Q=\frac{dM}{dr_+}/ \left(\frac{dT}{dr_+} \right)$ diverges. See bellow. It must be noticed, however, that for GR ($n=1$) in four dimensions ($d=4$) and for the second family of solutions with $n=2$ and $d=6$ the temperature is always an increasing function of $r_+$ and thus $r_*$ and $r_{**}$ disappear.  

\subsection{Heat Capacity}
The heat capacity can be defined for $Q$ constant as
\begin{equation}
  C_{Q} =\frac{dM}{dr_+} \left (\frac{dT}{dr_+} \right)^{-1} .
\end{equation}

\begin{figure}
\centering
\subfigure[ $C_Q$ for $n=1,d=4,5,6$.]{\includegraphics[width=80mm]{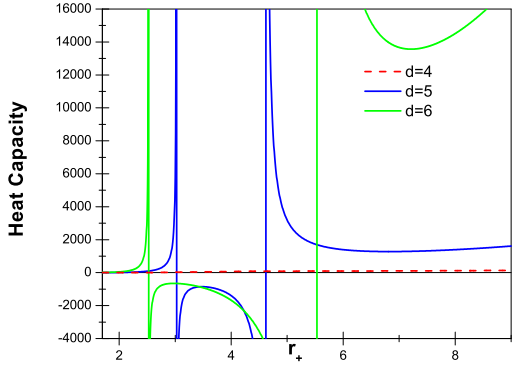}} 
\subfigure[ $C_Q$ for $n=3,d=8,9,10$.]{\includegraphics[width=80mm]{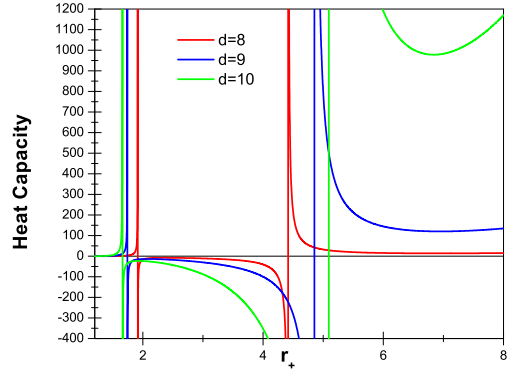}} 
\caption{$C_Q$. Figure a) represents both families of solutions, and figure b) represents to the first family of solutions, both with toy values of $Q=1$ and $l=8$ .}
\label{CalorEspecificoPL}
\end{figure}

\begin{figure}
\centering
\subfigure[$C_Q$ for $n=2,d=6,7,8$.]{\includegraphics[width=80mm]{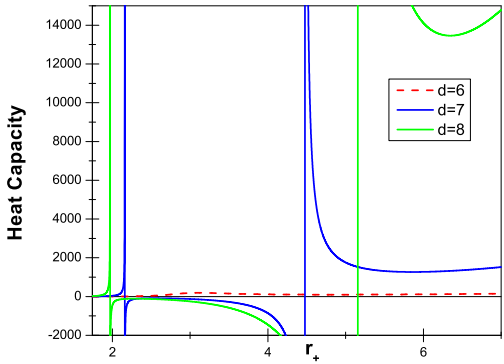}} 
\subfigure[$C_Q$ for $n=3,d=8,9,10$.]{\includegraphics[width=80mm]{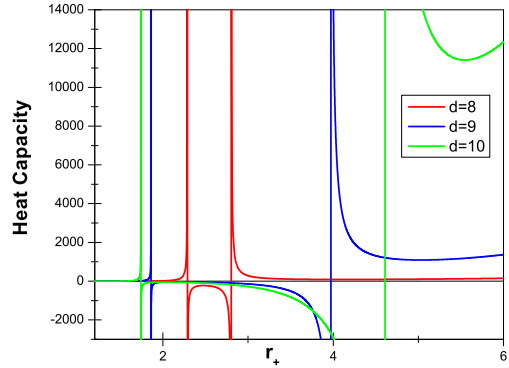}} 
\caption{$C_Q$ for second family of solutions with toy values of $Q=1$ and $l=8$ .}
\label{CalorEspecificoBHS}
\end{figure}

This definition, following the rules of thermodynamics, allows to determine that black holes are locally stable/unstable under thermal fluctuation provided $C_Q>0$ or $C_Q<0$ respectively. With this in mind one can proceed with the analysis of the solutions. Figures \ref{CalorEspecificoPL} and \ref{CalorEspecificoBHS} display examples of the behavior of the heat capacity.  Notice that Figure \ref{CalorEspecificoPL}(a) represents both families of solutions as $n=1$. 

In general, it can be observed a phase transitions at both $r_*$ and $r_{**}$. Going from $C_Q > 0$ for $r_+ > r_{**}$ to $C_Q < 0$ for $r_{*} < r_+ < r_{**}$. Finally $C_Q > 0$ for $r_+<r_{*}$. Moreover, one can notice as well that $C_{Q}$ vanishes as $T \rightarrow 0$. 

It is worth to mention that for GR ($n=1$) in four dimensions ($d=4$) and for the second family of solutions with $n=2$ and $d=6$ there is no phase transitions. Moreover, although cannot be observed at the figures, the specific heat in this case is a positive function of $r_+$ which vanishes as $T \rightarrow 0$ as well.

\section{Conclusions and Discussions}

In this work two families of regular black hole solutions have been discussed. Although each family is a solution of a different Lovelock gravity in $d$ dimensions both share to have a single ground state, which is approached asymptotically by the solutions and is defined by a single cosmological constant. These theories correspond to the {\it Pure Lovelock} theory \cite{14,Cai:2006pq,3} and to the theory discussed in \cite{Crisostomo:2000bb} which have a $n$-fold degenerated ground state. 

First it must be noticed that solutions have a minimum value of the parameter $M$, called $M_{cri}$ above, to represent a black hole geometry. As expected, these solutions asymptotically, says for large $r$, are indistinguishable from the previously known solutions \cite{Cai:2006pq,Crisostomo:2000bb} in vacuum. On the other hand, both families of solutions near the origin, for $r \sim 0$, behaves as maximally symmetric spaces which can be fixed, unambiguously, to be of positive curvature, {\it i.e.} the parameters can be fixed such as the solutions approach de Sitter spaces, as required to model a regular black hole, at their origins. 

Concerning the thermodynamics of the solutions it was shown that, although both families of solutions differ, in general terms their thermodynamics presents the same features.  The associate temperatures of the horizons have, in general, a local maximum and a local minimum at $r_*$ and $r_{**}$ respectively. Indeed, one can notice that there can be three different values of $r_+$ defining the same temperature. Because of this last, the heat capacity changes from positive for $r_+ < r_{*} $ to negative at $r_* < r_+ < r_{**}$ and finally to positive for $r_{**} < r_+$, signing out two phase transitions of the system, from thermodynamically stable $(C_Q >0)$ into unstable $(C_Q <0)$ and  viceversa. Therefore, in general, there is a range of the black hole radii, $ r_{*} < r_+ < r_{**}$, or equivalently of the masses, where the black holes become thermodynamically unstable. The existence of a phase transition for the vacuum solutions in \cite{Crisostomo:2000bb} was known, and thus the phase transition at $r_{**}$ could have been anticipated. However the existence of a second phase transition at $r_+ = r_{*}$, and thus the existence a second stable range for $r_+ < r_{*}$, is a new feature proper of these regular black holes. 

In order to analyze the thermodynamics in detail it was considered as model the generalized Hayward energy density \cite{Hayward:2005gi} defined in Eq,(\ref{FiniteMass}). Now, as mentioned in section \ref{AWellPosedMassDensity}, the conditions to be satisfied by the mass density are quite general, and thus it can be quite interesting to explore other options in order to determine to what extent the arise of phase transitions is a model dependent feature. For instance, it would be quite interesting to study a generalization of the proposal in \cite{7,17} 

\begin{acknowledgments}
This work was partially funded by grants FONDECYT 1151107. R.A. likes to thank DPI20140115 for some financial support.
\end{acknowledgments}

\appendix

\section{Units}\label{Units}

The unit system used throughout the text is slightly different from the ones presented in other articles. Because of that the unit system will be reviewed here in some detail.  

To begin with it has been fixed $\hbar = 1$ and $c=1$, the light speed. Therefore, the mass/energy units is actually $L^{-1}$, where $L$ is some distant unit. Moreover, it must be noticed that since $\hbar = 1$ the action principle must be dimensionless. 

As the $\alpha_n$  coefficients must be fixed such that $\alpha_n L_n$, with $L_n$ the $n$ term in the Lovelock Lagrangian, be dimensionless, and $[\sqrt{g} d^d x] = L^d$ then $[\alpha_n] = L^{2n-d}$. By the same token, one must notice that the mass density must satisfy $[\rho(r)] = L^{-d}$. Finally, by direct observation, the regulator in the generalized Hayward density must satisfy that $[Q] = L^{d/(d-2)}$. One can check that the term,  $\alpha_n^{-1} r^{1+2n-d} m(r)$, presented in the $f(r)$ function for both solution, satisfies
\[  
\left[\alpha_n^{-1} \frac{m(r)}{r^{d-2n-1}} \right] = L^0,
\]
as expected for consistency. 

\bibliography{mybib}

\end{document}